\documentclass[12pt,a4paper,english]{revtex4}
\usepackage[T1]{fontenc}
\usepackage[latin1]{inputenc}
\usepackage{amsmath}
\usepackage{amssymb}
\usepackage{graphicx}
\usepackage{esint}

\makeatletter

\pdfpageheight\paperheight
\pdfpagewidth\paperwidth

\providecommand{\tabularnewline}{\\}

\@ifundefined{textcolor}{}
{%
 \definecolor{BLACK}{gray}{0}
 \definecolor{WHITE}{gray}{1}
 \definecolor{RED}{rgb}{1,0,0}
 \definecolor{GREEN}{rgb}{0,1,0}
 \definecolor{BLUE}{rgb}{0,0,1}
 \definecolor{CYAN}{cmyk}{1,0,0,0}
 \definecolor{MAGENTA}{cmyk}{0,1,0,0}
 \definecolor{YELLOW}{cmyk}{0,0,1,0}
}

\usepackage{babel}

\makeatother

\usepackage{babel}
\begin{document}

\title{Attosecond electronic and nuclear quantum photodynamics of ozone:
time-dependent Dyson orbitals and dipole}

\author{A. Perveaux,$^{1,2}$D. Lauvergnat,$^{2}$ B. Lasorne,$^{1}$ F.
Gatti$^{1}$, M. A. Robb,$^{3}$ G. J. Halász,$^{4}$and Á. Vibók$^{5}$}

\affiliation{$^{1}$CTMM, Institut Charles Gerhardt Montpellier, Université Montpellier
2, F-34095 Montpellier, France}

\affiliation{$^{2}$Laboratoire de Chimie Physique, Bâtiment 349, CNRS, UMR8000,
Orsay, F-91405; Univ Paris-Sud, Orsay, F-91405; France }

\affiliation{$^{3}$Imperial College London, Department of Chemistry, London SW7
2AZ, UK}

\affiliation{$^{1}$Department of Information Technology, University of Debrecen,
H-4010 Debrecen, PO Box 12, Hungary}

\affiliation{$^{4}$Department of Theoretical Physics, University of Debrecen,
H-4010 Debrecen, PO Box 5, Hungary}

\email{vibok@phys.unideb.hu}

\begin{abstract}
A nonadiabatic scheme for the description of the coupled electron
and nuclear motions in the ozone molecule was proposed recently (PRA,\textbf{88},023425,
(2013)). An initial coherent nonstationary state was prepared as a
superposition of the ground state and the excited Hartley band. In
this situation neither the electrons nor the nuclei are in a stationary
state. The multiconfiguration time dependent Hartree method was used
to solve the coupled nuclear quantum dynamics in the framework of
the adiabatic separation of the time-dependent Schrödinger equation.
The resulting wave packet shows an oscillation of the electron density
between the two chemical bonds. As a first step for probing the electronic
motion we computed the time-dependent molecular dipole and the Dyson
orbitals. The latter play an important role in the explanation of
the photoelectron angular distribution. Calculations of the Dyson
orbitals are presented both for the time-independent as well as the
time-dependent situations. We limited our description of the electronic
motion to the Franck-Condon region only due to the localization of
the nuclear wave packets around this point during the first 5\textendash{}6
fs.
\end{abstract}
\maketitle

\section{Introduction}

In recent years, essential effort has been made to develop different
attosecond techniques, see, e.g., \cite{Feri1,Feri2,Feri3} and further
references therein. These techniques are based on an appropriate construction
of single ultrashort pulses or trains of such pulses, which allows
one to take real-time snapshots of ultrafast transformations of matter.
Pump-probe experimental techniques using these ultra short laser pulses
\cite{Feri3,Murnane,Huis1,Smirnova,Murnane-a,Murnane-2a} have made
possible to control complex molecular processes. Experimentalists
can excite or ionize atoms with a controlled few-cycle laser field
and then probe them through the spectrally resolved absorption of
an attosecond XUV pulse. It is the simplest experimental technique
used to study ultrafast electronic dynamics. During this process one
is able to fully image the electronic quantum motion and determine
the degree of coherence in the studied system \cite{santra1,santra2,elefteriosz1,santra3,kelken1}.
However, the real challenge now is how to transfer this technique
to molecules. 

Despite the fact that the electron dynamics in molecules almost always
are strongly coupled to nuclear dynamics in case of diatomics with
only one and two electrons the proper description of the nuclear-electron
dynamics is satisfactorily solved. In this case the total time-dependent
Schrödinger equations (TDSE) can be solved numerically including both
the electronic and nuclear degrees of freedom explicitly \cite{mainz1aa,mainz2aa,vrakking1a,regine1a,regine2a,regine3a,regine4a,steffi2a,moshammer2a,fernando1a,fernando2a}.
Another branch of methods treat the electron dynamics very accurately,
while keeping fixed the nuclear geometry. This is a fairly reasonable
assumption if the masses of the nuclei are large. In this situation
the motion of the electrons takes place due to the pump of a nonstationary
electronic state during a period of time that is much shorter than
the period of the vibrational motion of the nuclei \cite{Levine1a,Levine2a,Levine2aa,Levine3a,Levine4a,Levine5a,Nest1a}.
The really challenging task is to create an electronic wave packet
in a neutral few-atom system and then describe precisely its coupled
electron nuclear dynamics. Such systems are really large enough not
to use the advantage of the solution of the TDSE without separation,
but they are not large enough to use the rigid nuclear geometry assumption.
In this case the difference between the nuclear vibration period and
the period of the motion of the electrons are not always very significant. 

Recently, we have investigated the ozone molecule and proposed a new
scheme for the description of the coupled electron and nuclear motion
\cite{Agnee1a,Agnee2a}. An initial coherent nonstationary state was
prepared by a sub-femtosecond pulse \cite{Agnee2a}. It is a superposition
of the ground state X and the excited weakly-bound state B of the
Hartley band \cite{Schinke2a,Schinke3a,Schinke1a,Gabriel}. In this
situation neither the electrons nor the nuclei are in a stationary
state, and we used quantum dynamics simulations. The nuclear wave
packets, the electronic populations, the relative electronic coherence
between the ground X and B electronic states and the electron wave
packet dynamics were analyzed. The time evolution of the electronic
motion was plotted in the Franck-Condon (FC) region only due to the
localization of the nuclear wave packet around this point during the
first $5-6$ fs. 

The purpose of this paper is to go further in the photodynamics simulation
of this system. We now present the time-dependent dipole moment \cite{Vrakking1}
and investigate the Dyson orbitals \cite{Krylov1,Krylov2,Remacle1,serguei1,spanner}.
While the former one is useful to visualize the exciton migration
as the electron density shows a fast oscillation between the two chemical
bonds, the Dyson orbitals are known to be central in the explanation
of the photoelectron angular distribution. Both the time-independent
as well as the time-dependent Dyson orbitals are calculated and their
properties are discussed.

In Sec. II. we give a short description of the chosen theoretical
approaches. The relevant expressions and formalism will also be presented
here. The results and the discussions will be presented in Sec. III.
Section IV. provides conclusions.

\section{Methods and Formalism}

The adiabatic partition formalism (beyond Born-Oppenheimer \cite{Baer})
assumes the total molecular wave function $\Psi_{tot}(\vec{r}_{el},\vec{R},t)$
as a sum of products of electronic wave functions, $\psi_{el}^{k}(\vec{r}_{el};\vec{R})$,
and nuclear wave packets, $\Psi_{nuc}^{k}(\vec{R},t)$:

\begin{equation}
\Psi_{tot}(\vec{r}_{el},\vec{R},t)=\sum_{k=1}^{n}\Psi_{nuc}^{k}(\vec{R},t)\psi_{el}^{k}(\vec{r}_{el};\vec{R}).\label{eq:totwave}
\end{equation}

Here $k$ denotes the $k-th$ adiabatic electronic state, $\vec{r}_{el}$
and $\vec{R}$ are the electronic and the nuclear coordinates, respectively.
We are interested in solving the coupled evolution of the nuclear
wave packets, $\Psi_{nuc}^{k}(\vec{R},t)$, by inserting the product
ansatz (\ref{eq:totwave}) into the time-dependent Schrödinger equation
of the full molecular Hamiltonian. The electronic wave function obeys
the time-independent Schrödinger equation for the electronic Hamiltonian
$H_{el}(\vec{R})$

\begin{equation}
H_{el}(\vec{R})\psi_{el}^{l}(\vec{r}_{el};\vec{R})=V_{l}(\vec{R})\psi_{el}^{l}(\vec{r}_{el};\vec{R}),\label{eq:elect-sch}
\end{equation}
and integrating over the electronic coordinates we can also obtain
the coupled nuclear Schrödinger equations:

\begin{equation}
i\hbar\frac{\partial}{\partial t}\Psi_{nuc}^{k}(\vec{R},t)=\sum_{l=1,n}H_{k,l}(\vec{R})\Psi_{nuc}^{l}(\vec{R},t).\label{eq:nuc-sch}
\end{equation}
Here $H_{k,l}$ is the matrix element of the vibronic Hamiltonian,
which reads, e.g., for $n=2$,

\begin{equation}
H=\left(\begin{array}{cc}
T_{nuc}+V_{k} & K_{k,l}^{*}\\
K_{k,l} & T_{nuc}+V_{l}
\end{array}\right),\label{eq:nuc-ham}
\end{equation}
where $T_{nuc}$ is the nuclear kinetic energy, $V_{k}$ ($k=1,...n$)
is the $k-th$ adiabatic potential energy and $K_{k,l}$ with $k\neq l$
is the coupling term between the $(k,l)-th$ electronic states, which
contains the light-matter interaction, $-\vec{\mu}_{k,l}\cdot\overrightarrow{E}(t)$
(electric dipole approximation), where $\vec{E}(t)$ is an external
field resonant between the $k-th$ and the $l-th$ states and $\vec{\mu}_{k,l}$
is the $\vec{R}-$dependent transition dipole moment. 

One has to solve the electronic and nuclear Schrödinger equations
Eqs. (\ref{eq:elect-sch}-\ref{eq:nuc-sch}) to obtain the potential
energy surfaces and the electronic and nuclear wave functions. The
electronic structure calculations are performed by the MOLPRO code
\cite{molpro-1} and a development version of the GAUSSIAN program
package \cite{gaussian}. As for the nuclear dynamics calculations
the multiconfiguration time-dependent Hartree method (MCTDH) method
\cite{dieter1a,dieter2a,dieter3a,dieter4a} was used.

The MCTDH nuclear wave packets, $\Psi_{nuc}^{k}(\vec{R},t)$, contain
all the information about the relative phases between the electronic
states. Using the interaction picture, $\Psi_{nuc}^{k}(\vec{R},t)$
can equally be written as: 
\begin{equation}
\Psi_{nuc}^{k}(\vec{R},t)=\exp(-iV_{k}(\vec{R})t/\hbar)a_{k}(\vec{R},t).\label{eq:phase}
\end{equation}
Here, $V_{k}(\vec{R})$ is the potential energy of the $k-th$ state.
The first part of this wave function is the phase factor, ($\exp(-iV_{k}(\vec{R})t/\hbar)$),
of the $k-th$ state, which oscillates very fast. 

From the electronic and nuclear wave functions the total density matrix
of the molecule, the electronic populations on the different states
and the electronic relative coherence between the different electronic
states have been calculated \cite{Agnee1a,Agnee2a}. 

In the present case we have two states (the ground X and the Hartley
B states), and the electronic wave packet of the neutral molecule
can be written as follows

\begin{equation}
\Psi_{tot}(\vec{r}_{el},\vec{R},t)=\Psi_{nuc}^{X}(\vec{R},t)\psi_{el}^{X}(\vec{r}_{el};\vec{R})+\Psi_{nuc}^{B}(\vec{R},t)\psi_{el}^{B}(\vec{r}_{el};\vec{R}).\label{eq:2-states-wave}
\end{equation}

Applying this formula, the total time-dependent dipole of the molecule
reads as

\begin{equation}
\vec{\mu}{}_{tot}(\vec{R},t)=\left\langle \Psi_{tot}(\vec{R},t)\mid\vec{\mu}{}_{tot}\mid\Psi_{tot}(\vec{R},t)\right\rangle =\sum_{k,l=X,B}\Psi_{nuc}^{k*}(\vec{R},t)\Psi_{nuc}^{l}(\vec{R},t)\vec{\mu}_{k,l}(\vec{R})\label{eq:time-dep-dipol}
\end{equation}
here $\vec{\mu}_{k,l}$ is the transition dipole moment between the
$k-th$ and $l-th$ electronic states. 

Probing the electronic wave packet will lead to ionize the ozone molecule.
The Dyson orbitals correspond to the molecular orbitals of the neutral
molecule from which an electron has been removed where the cation
relaxation is accounted for. They can be computed as one-electron
transition amplitudes between the N-electron neutral and (N-1) - electron
cationic states:

\begin{equation}
\Phi_{cat}^{D}(\vec{r};\vec{R})=\sqrt{N}\int d\vec{r}_{1}...d\vec{r}_{N-1}\psi_{el,neut}^{N}(\vec{r}_{1},...\vec{r}_{N}=\vec{r};\vec{R})\psi_{el,cat}^{N-1}(\vec{r}_{1},...\vec{r}_{N-1};\vec{R}).\label{eq:dyson1}
\end{equation}

Applying the occupation number representation the Dyson orbitals can
also be expressed in the molecular orbitals of the neutral molecule 

\begin{equation}
\Phi_{cat}^{D}(\vec{r};\vec{R})=\sum_{k}\varphi_{k}^{neut}(\vec{r})\left\langle \psi_{el,cat}^{N-1}(\vec{R})\mid\hat{a}_{k}\psi_{el,neut}^{N}(\vec{R})\right\rangle .\label{eq:dyson2}
\end{equation}

Here $\hat{a}_{k}$ is the operator which removes an electron from
the molecular orbital $\varphi_{k}$ . 

We now define the time-dependent Dyson orbitals. These orbitals may
be useful for instance when the neutral molecule is excited by an
ultrashort laser pulse creating a coherent superposition of the different
stationary states in the neutral molecule that will be probed in the
next step by sudden XUV ionization. We focus here on such a situation. 

\begin{align}
\Phi_{cat,i}^{D}(\vec{r};\vec{R},\tau) & =\sqrt{N}\int d\vec{r}_{1}...d\vec{r}_{N-1}\psi_{el,neut}^{N}(\vec{r}_{1},...\vec{r}_{N}=\vec{r};\vec{R},\tau)\psi_{el,cat,i}^{N-1}(\vec{r}_{1},...\vec{r}_{N-1};\vec{R})\nonumber \\
= & \sum_{k}\Psi_{nuc}^{k}(\vec{R},\tau)\Phi_{cat,i}^{D}(\vec{r};\vec{R}),\label{eq:time-dependent-dyson}
\end{align}
here $\tau$ is the time when the ionization takes place. $i$ denotes
the different cation channels. At a given $\vec{R}$, $\psi_{el,neut}^{N}(\vec{r}_{1},...\vec{r}_{N};\vec{R},\tau)$
is the electronic wave packet, which is a coherent superposition of
the ground (X) and the Hartley (B) states, and reads as

\begin{equation}
\psi_{el,neut}^{N}(\vec{r}_{el};\vec{R},\tau)=\Psi_{tot}(\vec{r}_{el},\vec{R},\tau)=\Psi_{nuc}^{X}(\vec{R},\tau)\psi_{el}^{X}(\vec{r}_{el};\vec{R})+\Psi_{nuc}^{B}(\vec{R},\tau)\psi_{el}^{B}(\vec{r}_{el};\vec{R}).\label{eq:elect-wave-packet}
\end{equation}
From this quantity one can form the time-dependent density of the
Dyson orbitals at the FC geometry

\begin{equation}
\rho_{\Phi_{cat}^{D}}(\vec{r};\vec{R},\tau)=\sum_{k,l=X,B}\Psi_{nuc}^{k*}(\vec{R},\tau)\Psi_{nuc}^{l}(\vec{R},\tau)\Phi_{cat}^{D,k*}(\vec{r};\vec{R})\Phi_{cat}^{D,l}(\vec{r};\vec{R}).\label{eq:density-dyson}
\end{equation}
This density can be written for each cation channel as well

\begin{align}
\rho_{\Phi_{cat,i}^{D}}(\vec{r};\vec{R},\tau) & =\mid\Psi_{nuc}^{X}(\vec{R},\tau)\mid^{2}\rho_{\Phi_{cat,i}^{D,X}}(\vec{r};\vec{R})+\mid\Psi_{nuc}^{B}(\vec{R},\tau)\mid^{2}\rho_{\Phi_{cat,i}^{D,B}}(\vec{r};\vec{R})\nonumber \\
+ & 2Re\Psi_{nuc}^{X*}(\vec{R},\tau)\Psi_{nuc}^{B}(\vec{R},\tau)\Phi_{cat,i}^{D,X*}(\vec{r};\vec{R})\Phi_{cat,i}^{D,B}(\vec{r};\vec{R})\label{eq:density-dyson-cat-channel}
\end{align}

where index $i$ runs over each of cation channel.

Let us define the local norm of the Dyson orbitals at the FC geometry,
which is approximately proportional to the ionization probability
(similar to a Franck-Condon factor in the impulsive picture)

\begin{equation}
\left\langle \Phi_{cat}^{D}(\vec{r};\vec{R},\tau)\mid\Phi_{cat}^{D}(\vec{r};\vec{R},\tau)\right\rangle =\sum_{k,l=X,B}\Psi_{nuc}^{k*}(\vec{R},\tau)\Psi_{nuc}^{l}(\vec{R},\tau)\left\langle \Phi_{cat}^{D,k}(\vec{r};\vec{R},\tau)\mid\Phi_{cat}^{D,l}(\vec{r};\vec{R},\tau\right\rangle .\label{eq:local-norm}
\end{equation}

For a certain cation channel this quantity reads as:

\begin{align}
\left\langle \Phi_{cat,i}^{D}(\vec{r};\vec{R},\tau)\mid\Phi_{cat,i}^{D}(\vec{r};\vec{R},\tau)\right\rangle  & =\mid\Psi_{nuc}^{X}(\vec{R},\tau)\mid^{2}\left\langle \Phi_{cat,i}^{D,X}(\vec{r};\vec{R},\tau)\mid\Phi_{cat,i}^{D,X}(\vec{r};\vec{R},\tau\right\rangle \nonumber \\
 & +\mid\Psi_{nuc}^{B}(\vec{R},\tau)\mid^{2}\left\langle \Phi_{cat,i}^{D,B}(\vec{r};\vec{R},\tau)\mid\Phi_{cat,i}^{D,B}(\vec{r};\vec{R},\tau\right\rangle \nonumber \\
+ & 2Re\Psi_{nuc}^{X*}(\vec{R},\tau)\Psi_{nuc}^{B}(\vec{R},\tau)\left\langle \Phi_{cat,i}^{D,X*}(\vec{r};\vec{R})\mid\Phi_{cat,i}^{D,B}(\vec{r};\vec{R})\right\rangle .\label{eq:local-norm-cation-chanel}
\end{align}

These expressions and equations will serve as our working formulae
in the next section.

\section{Results and Discussion}

Figure 1 shows the total time-dependent dipole Eq. (\ref{eq:time-dep-dipol})
of the molecule. In the present situation the time-dependent dipole
is created after the interaction of the molecule with the pump laser
field \cite{Vrakking1}. Its oscillatory behaviour can be considered
as a direct consequence of the coherent superposition of the X and
B electronic states. This quantity has a similar time evolution to
the electronic relative coherence, which illustrates exciton migration.
The FC point of the ozone molecule has $C_{2v}$ symmetry and therefore
only the \emph{y}-component ($B_{2}$) of the transition dipole between
the ground state X ($^{1}A_{1}$) and Hartley B ($^{1}B_{2}$) is
nonzero. The only effective polarization of the electric field is
$y$. 

\begin{figure}
\begin{centering}
\includegraphics[scale=0.7]{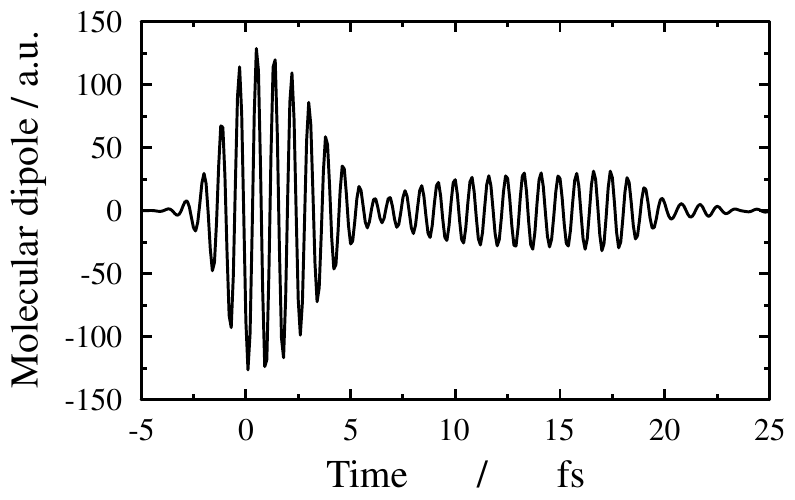}
\par\end{centering}

\caption{Time-dependent molecular dipole along the y-axis. }

\end{figure}

Probing the electronic wave packet Eq. (\ref{eq:elect-wave-packet})
will lead to ionize the neutral molecule. We focus here on the ionization
threshold of the spectrum (fastest electrons). The three lowest cationic
states ($^{2}A_{1}$$,^{2}B_{2}$, and $^{2}A_{2}$ ) must be considered
together because they are energetically close with respect to the
expected bandwith of the probe pulse. We have calculated the energies
of the different electronic states of the cation at the MRCI level
of theory and compared them to the experimental values \cite{experiment}.
Reasonably good agreement has been obtained for each electronic state.
Results are presented in Table 1.

\begin{table}
\begin{tabular}{|c|c|c|}
\hline 
Symmetry of the cation state & Excitation energy at the MRCI level {[}eV{]} & Experimental value {[}eV{]} \tabularnewline
\hline 
\hline 
$^{2}A_{1}$  & 12.28 & 12.73\tabularnewline
\hline 
$^{2}B_{2}$ & 12.41 & 13.00\tabularnewline
\hline 
$^{2}A_{2}$ & 13.09 & 13.54\tabularnewline
\hline 
\end{tabular}

\caption{\label{tab:1}Excitation energies calculated in aug-cc-pvqz basis
at the MRCI level of theory for each cation channel. Experimental
excitation energies are also presented \cite{experiment}.}
\end{table}

We now present the Dyson orbitals. They are very illustrative as they
are one-electron wave functions that represent the probability amplitude
of the electron that is removed from the neutral molecule. They also
play an important role in the interpretation of the photoelectron
angular distribution as well. Namely, they are approximately proportional
to the latter, as their norms give some contributions to the photoionization
intensity. In the present situation we have three ionization channels
for the ground (X) and also for the Hartley (B) states of the neutral
molecule. These six different time-independent Dyson orbitals are
calculated according to Eqs. (\ref{eq:dyson1} - \ref{eq:dyson2})
and are shown in Figure 2. The MRCI method with aug-cc-pvqz basis
set was used in the numerical simulations. 

\begin{figure}
\begin{centering}
\includegraphics[scale=0.5]{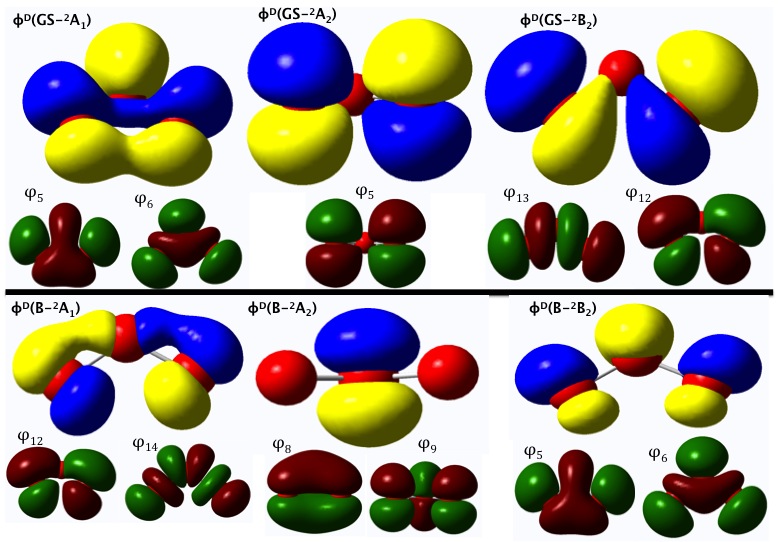}
\par\end{centering}

\caption{Time-independent Dyson orbitals between the two different electronic
states (X, B) of the neutral and the three different channels ($^{2}A_{1}$,
$^{2}B_{2}$ and $^{2}A_{2}$) of the cation. The most important molecular
orbital components of the Dyson orbitals are also shown.}

\end{figure}

As already discussed the Dyson orbitals can also be written as linear
combinations of the molecular orbitals of the neutral molecule Eq.
(\ref{eq:dyson1}). As it can be seen in Fig. 2 they follow the Koopman's
like rule. Hence, the Dyson orbitals for each cation channel are constructed
mainly from one or two neutral's orbitals. Below we give the most
important molecular orbitals contributions to the different type of
time-independent Dyson orbitals calculated for the present situation:

\begin{align}
({}^{2}A_{1})\Phi_{X-^{2}A_{1}}^{D} & =-0.61\varphi_{5}+0.45\varphi_{6};\nonumber \\
({}^{2}A_{2})\Phi_{X-^{2}A_{2}}^{D} & =-0.79\varphi_{15};\nonumber \\
({}^{2}B_{2})\Phi_{X-^{2}B_{2}}^{D} & =0.62\varphi_{12}-0.24\varphi_{13};\label{dy-mol-comp}\\
({}^{2}B_{2})\Phi_{B-^{2}A_{1}}^{D} & =-0.22\varphi_{12}-0.062\varphi_{14};\nonumber \\
({}^{2}B_{1})\Phi_{B-^{2}A_{2}}^{D} & =0.16\varphi_{8}-0.073\varphi_{9};\nonumber \\
({}^{2}A_{1})\Phi_{B-^{2}B_{2}}^{D} & =-0.24\varphi_{5}+0.0993\varphi_{6}.\nonumber 
\end{align}

The symmetries of the Dyson orbitals are provided by the symmetries
of the cation states and the symmetries of the X and B states of the
neutral. 

Our initial wave function of the neutral ozone molecule is a wave
packet, Eq. (\ref{eq:2-states-wave}), which is a superposition of
two electronic states (X and B). Therefore, the Dyson orbitals are
the superpositions of Dyson orbitals from each electronic states of
the neutral to the cation. Thus, we are going to have three time-dependent
Dyson orbitals Eq. (\ref{eq:time-dependent-dyson}); one for each
cation channel ($^{2}A_{1}$, $^{2}B_{2}$ and $^{2}A_{2}$). 

In Figures (3-5) the time-dependent densities of the Dyson orbitals
are presented. They are calculated according to Eqs. (\ref{eq:time-dependent-dyson},\ref{eq:density-dyson},\ref{eq:density-dyson-cat-channel}).
It can be noticed that the time-dependent densities of the Dyson orbitals
are quite similar to the graphical representation of the time-independent
Dyson orbitals between the X state of the neutral and an appropriate
electronic state of the cation. This is due to the fact that the population
on the ground state is always more pronounced than in the excited
state (in our case it is the Hartley (B) band). The time-dependent
densities of the Dyson orbitals oscillate in time with the same period
as the dipole moment of the neutral wave packet which is the period
of the coherence. Therefore, there are three different stages in the
time evolution of the time-dependent densities of the Dyson orbitals.
The first period is between {[}$-5.5$ ; $6${]} fs when the laser
light is on and there is significant relative electronic coherence
between the X and B states. In the second period {[}$5.5$ ; $8${]}
fs the laser pulse is off and there is no coherence between the different
electronic states. Finally, the third period is {[}$9$ ; $20${]}
fs when the pump pulse is still off but the electronic coherence reappears
between the ground X and the Hartley B states of the ozone molecule.
The oscillations of the time-dependent densities are the most pronounced
in the first interval as the relative electronic coherence is the
most significant here. In the second region there is practically no
motion of the time-dependent densities due to a lack of coherence.
As for the third stage, the revival of electronic coherence induces
a revival of motion in the time-dependent densities but with much
less amplitude than the first interval. The latter is due to the fact
that the coherence in the field free situation is less prominent than
in presence of the light field. Although all three regions were properly
included in the numerical simulations the time evolution of the time-dependent
densities in the snapshots (Figs. (3-5)) are only presented between
the period of {[}$-0.3$ ; $10.4${]} fs. 

\begin{figure}
\begin{centering}
\includegraphics[scale=0.5]{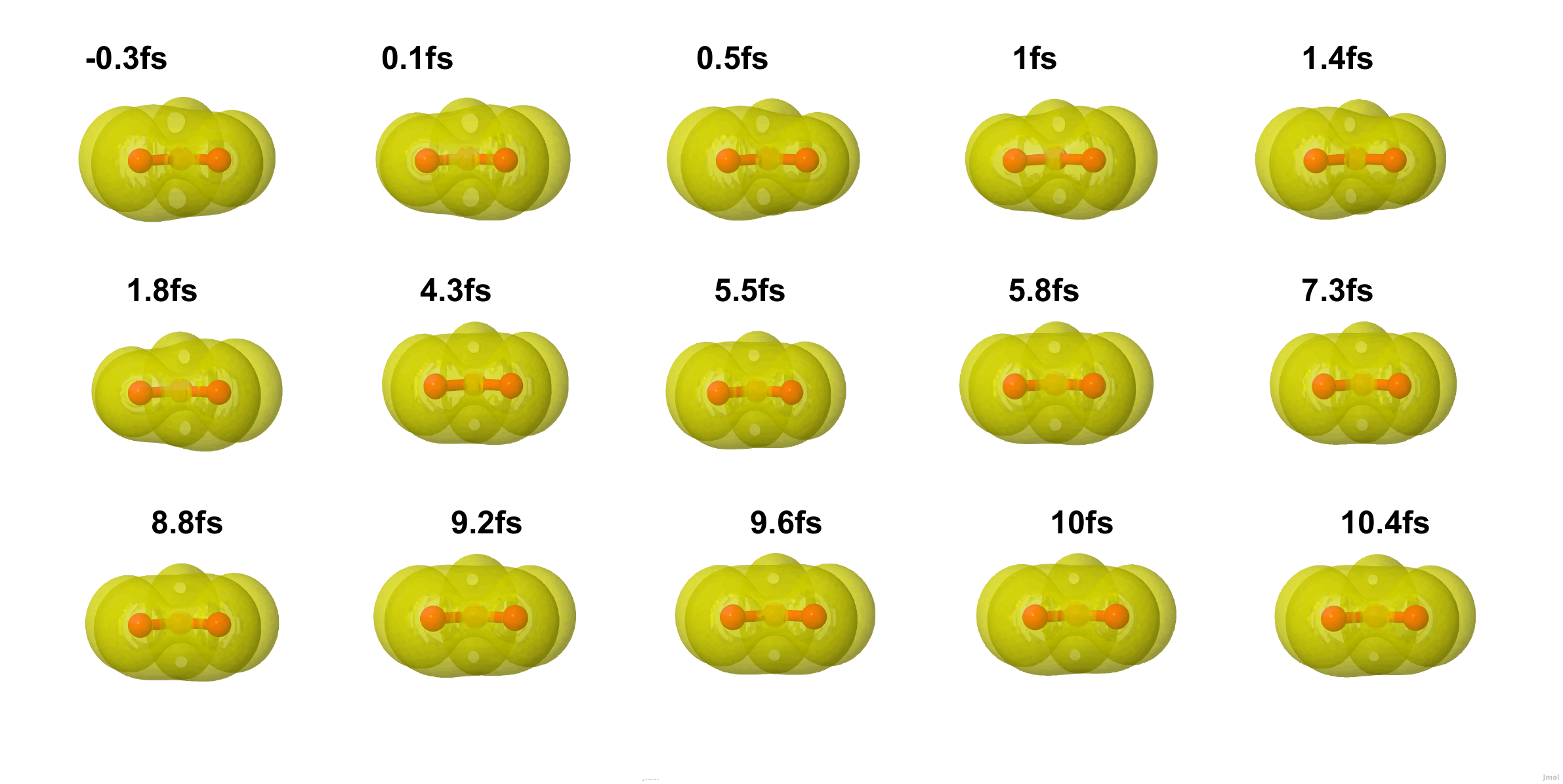}
\par\end{centering}

\caption{Time-dependent densities of the Dyson orbitals belong to $^{2}A_{1}$
cation channel.}
\end{figure}

\begin{figure}
\begin{centering}
\includegraphics[scale=0.5]{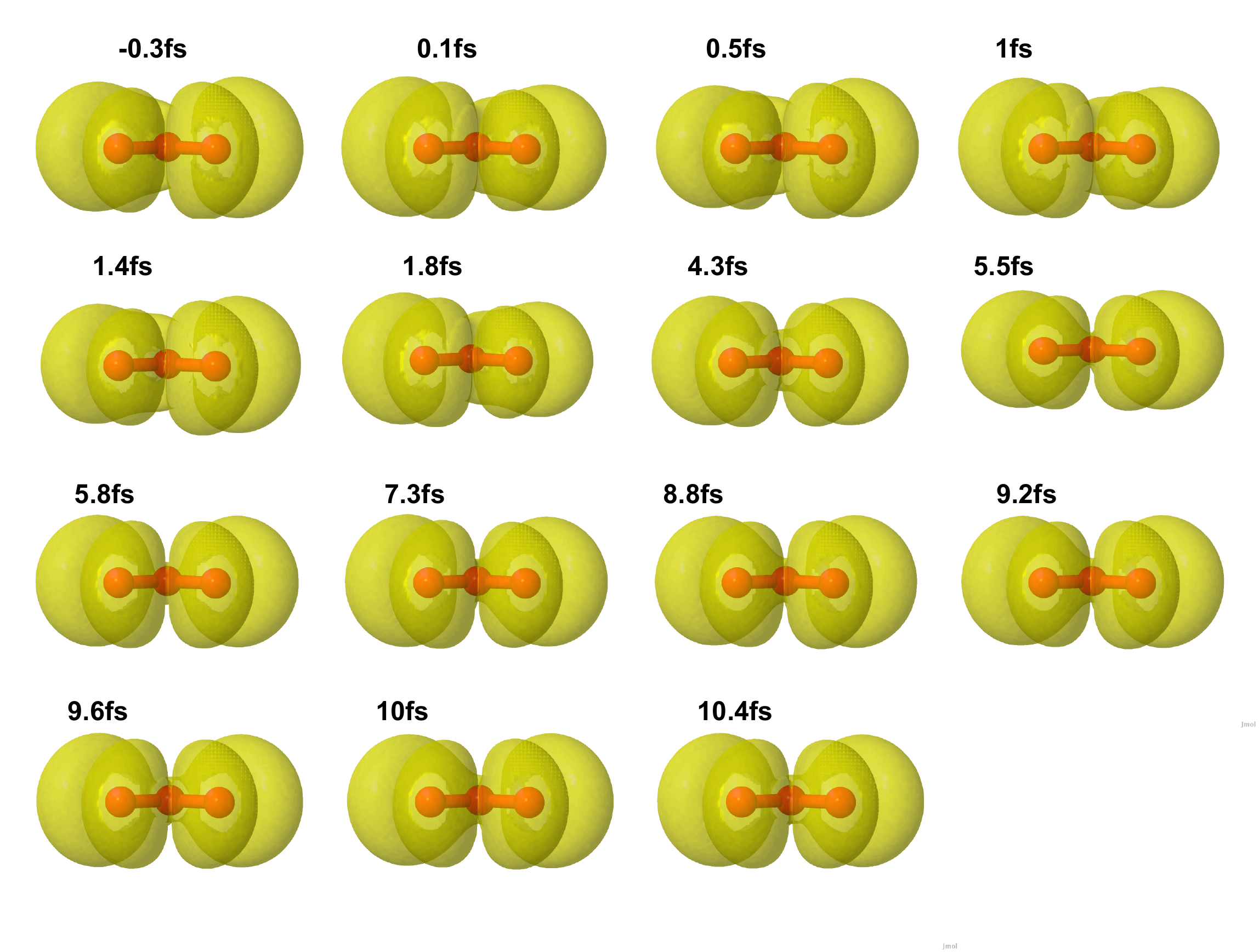}
\par\end{centering}

\caption{Time-dependent densities of the Dyson orbitals belong to $^{2}B_{2}$
cation channel.}

\end{figure}

\begin{figure}
\begin{centering}
\includegraphics[scale=0.5]{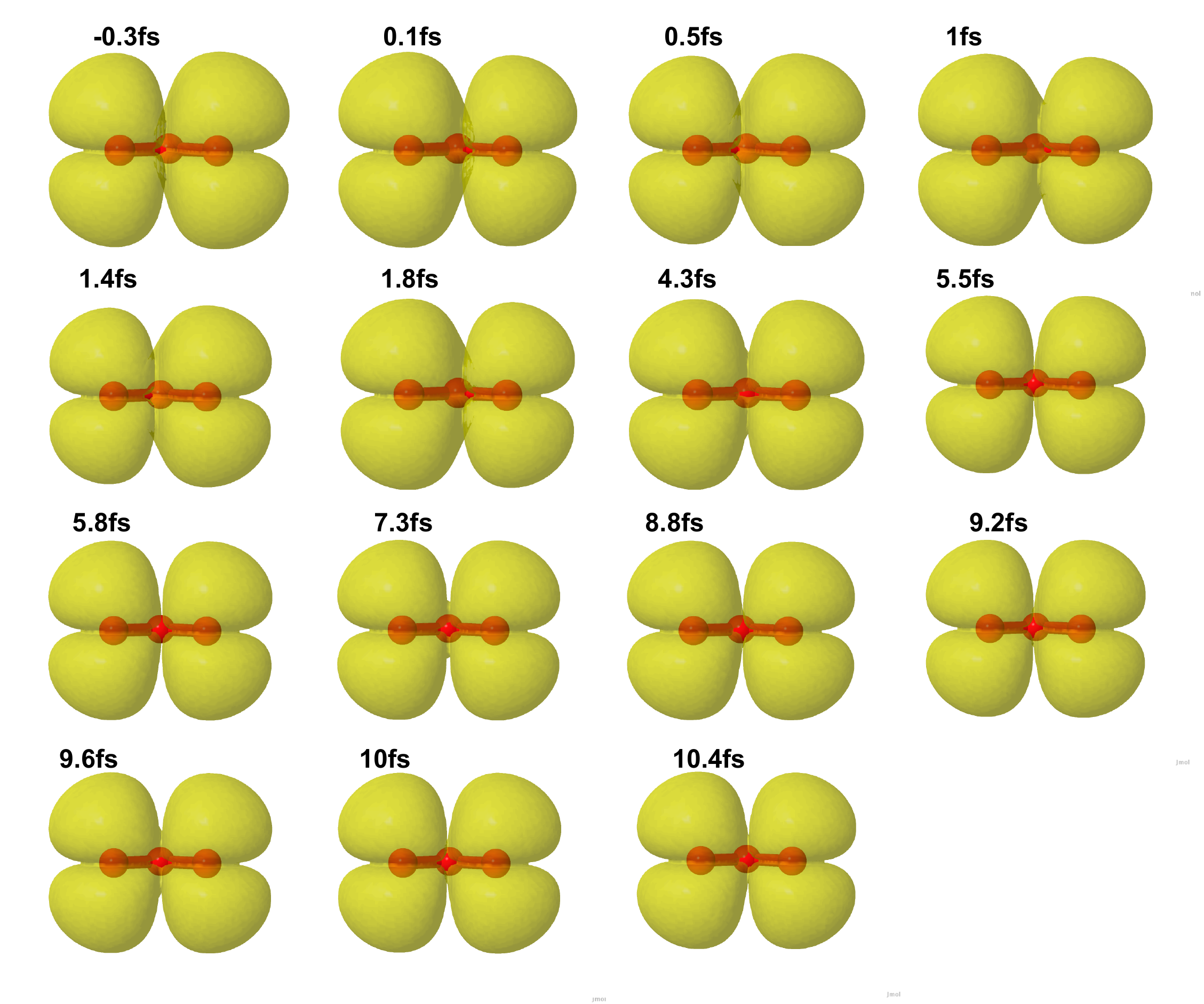}
\par\end{centering}

\caption{Time-dependent densities of the Dyson orbitals belong to $^{2}A_{2}$
cation channel.}

\end{figure}

We have also calculated the local norms of the Dyson orbitals Eqs.
(\ref{eq:local-norm},\ref{eq:local-norm-cation-chanel}). These quantities
are approximately proportional to the ionization probabilities. Probability
densities at the FC geometry, are shown in Figure 6. They can be considered
as \textquotedbl{}local norms\textquotedbl{}, or rather local weights,
with values higher than 1. At first sight, the striking feature on
the pictures (see Figs. 6a - 6b) is that the ground X state is the
essential component of the Dyson orbital's norm. This is consistent
with the previous finding that the population in the ground X state
is more pronounced than that one on the Hartley B state. We can also
notice, that the cation channel that presents more coherence ($^{2}B_{2}$)
is at the same time the one that provides less total probability.
And the reverse is also true, namely that the one presenting less
coherence ($^{2}A_{2}$) is the one presenting more total probability.
It is partly explained by the fact that the more there is population
on the ground X state, the less there is on the Hartley B state, therefore
the less there is electronic coherence between these two (neutral
and cation) excited states (see Fig. 6c). On the other hand the more
there is population on the ground X state the more the probability
of the time-dependent Dyson orbitals increases.

\begin{figure}
\begin{centering}
\includegraphics[width=0.45\textwidth]{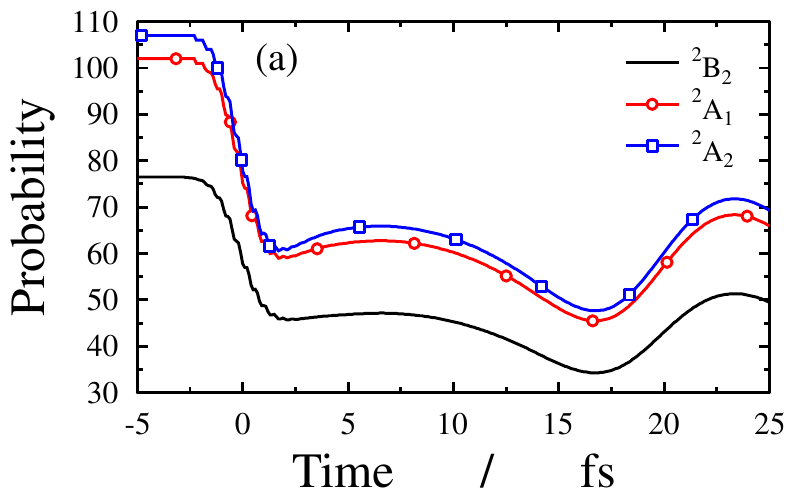}\includegraphics[width=0.45\textwidth]{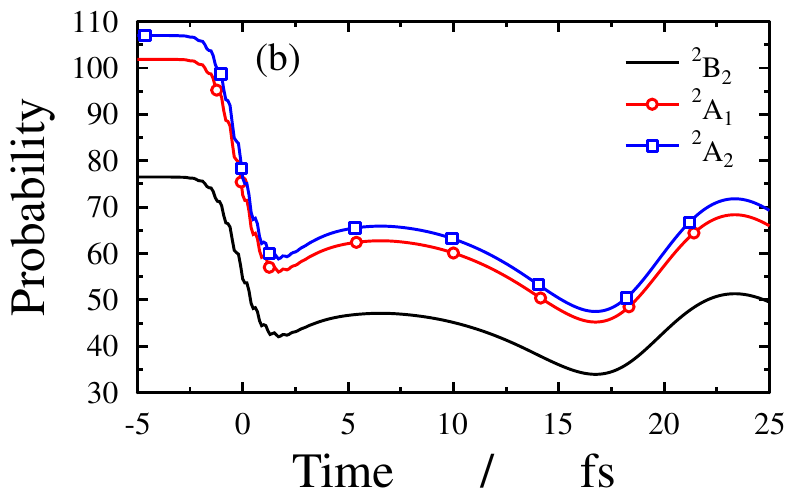}
\par\end{centering}

\begin{centering}
\includegraphics[width=0.45\textwidth]{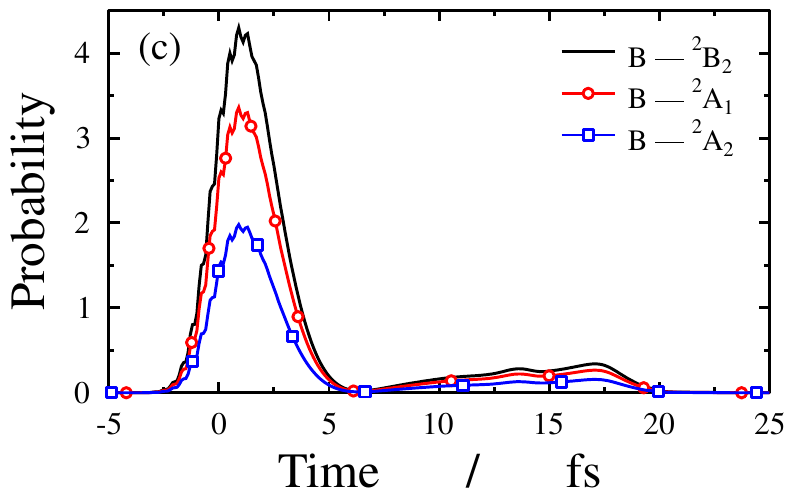}\includegraphics[width=0.45\textwidth]{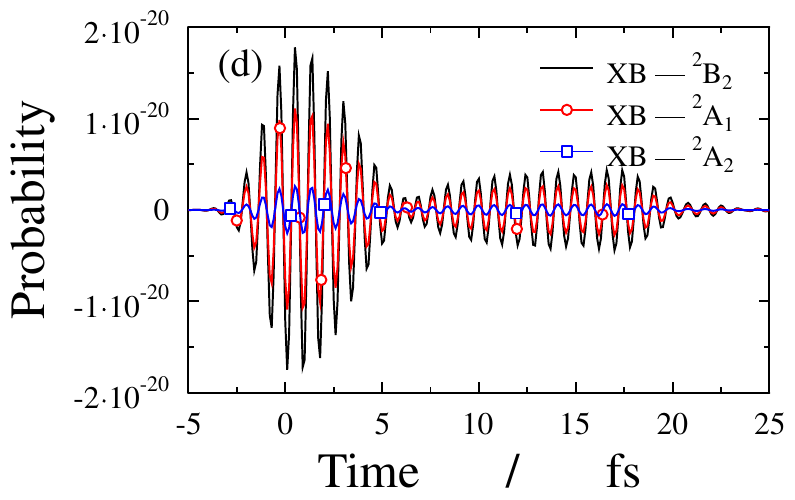}\caption{Probability density of the time-dependent Dyson orbitals between the
different states of the neutral molecule and the cation as a function
of time {[}fs{]}. (a): Total probability density of the time-dependent
Dyson orbitals as a function of time; $^{2}B_{2}$ channel (unmarked);
$^{2}A_{2}$ channel (marked with square); $^{2}A_{1}$ channel (marked
with circle). The total probability density can be obtained as a sum
of the probability densities on sub figures (b) and (c). (b): Probability
density of the time-dependent Dyson orbitals as a function of time
between the X state of the neutral and the different states of the
cation;$^{2}B_{2}$ channel (unmarked), $^{2}A_{2}$ channel (marked
with square); $^{2}A_{1}$ channel (marked with circle). (c): Probability
density of the time-dependent Dyson orbitals as a function of time
between the B state of the neutral and the different states of the
cation;$^{2}B_{2}$ channel (unmarked), $^{2}A_{2}$ channel (marked
with square); $^{2}A_{1}$ channel (marked with circle). (d): Probability
density of the relative electronic coherence of the X and B states
of the neutral with the $^{2}B_{2}$ cation channel (unmarked), $^{2}A_{2}$
cation channel (marked with square); $^{2}A_{1}$ cation channel (marked
with circle) as a function of time.}

\par\end{centering}

\end{figure}

\section{Conclusions}

We have started to develop a complex theoretical description of the
coupled electron and nuclear motion in the ozone molecule on the attosecond
time scale \cite{Agnee1a,Agnee2a}. An initial coherent nonstationary
state was created as a coherent superposition of the ground X and
excited Hartley B states. In this situation we were able to induce
attosecond electron dynamics in the neutral molecule. The MCTDH approach
was used to solve the dynamical Schrödinger equation for the nuclei
in the framework of the time-dependent adiabatic partition including
the light-matter interaction (electric dipole approximation). Based
on this dynamical simulation the description of the time evolution
of the electronic motion is limited only to the Franck-Condon region
due to the localization of the nuclear wave packet around this point
during the first $5-6$ fs. 

Applying the nuclear wave packet we have determined the total density
matrix of the molecule and from it were able to calculate the electronic
populations and the relative electronic coherence between the ground
X and B electronic states \cite{Agnee2a}. In order to calculate the
excited-state differential charge density at the FC point we used
the total molecular wave packet which is a coherent mixture of multiple
electronic states, whereby the time-dependent coefficients are the
nuclear wave packets. As a results, an oscillation of the electronic
cloud between the two chemical bonds was observed with a $0.8$ fs
period of time \cite{Agnee2a}. 

Going further in our photodynamics description we calculated the time-dependent
dipole moment and the Dyson orbitals. They are very useful to visualize
the exciton migration within the molecule as the electron density
oscillates between the two chemical bonds. In addition, the Dyson
orbitals are known to be important in the explanation of the photoelectron
angular distribution. We calculated both the time-independent as well
as the time-dependent Dyson orbitals and discussed their properties
for different situations. Corresponding experiments are in progress
\cite{Reinhard}.

\section*{Acknowledgements}

The authors would like to thank R. Kienberger , M. Jobst and F. Krausz,
for support and for fruitful discussions. We acknowledge R. Schinke
for providing the potential energy surfaces and the transition dipole
moment and H.-D. Meyer for fruitful discussions. The authors also
acknowledge the TÁMOP 4.2.4. A/2-11-1-2012-0001 project. Á.V. acknowledges
the OTKA (NN103251). Financial support by the CNRS-MTA is greatfully
acknowledged.

\end{document}